\documentclass[twocolumn]{aastex631}

\usepackage{CJKutf8}
\usepackage{amsmath}
\usepackage{xcolor}

\newcommand{\T}{\ensuremath{\mathrm{T}}}

\newcommand{\appropto}{\mathrel{\vcenter{
  \offinterlineskip\halign{\hfil$##$\cr
    \propto\cr\noalign{\kern2pt}\sim\cr\noalign{\kern-2pt}}}}}

\newcommand{\responseB}[1]{{#1}}

\graphicspath{{./}{figures/}}

\begin{document}

\title{Unexpected Near-Resonant and Metastable States of Young Multi-Planet Systems}

\author[0009-0000-6461-5256]{Zhecheng Hu (\begin{CJK*}{UTF8}{gbsn}胡哲程\end{CJK*})}
\affiliation{Department of Astronomy, Tsinghua University, Beijing 10084, China}
\affiliation{Institute for Astronomy, University of Hawai‘i, 2680 Woodlawn Drive, Honolulu, HI 96822, USA}

\author[0000-0002-8958-0683]{Fei Dai (\begin{CJK*}{UTF8}{gbsn}戴飞\end{CJK*})}
\affiliation{Institute for Astronomy, University of Hawai‘i, 2680 Woodlawn Drive, Honolulu, HI 96822, USA}

\author[0000-0003-4027-4711]{Wei Zhu (\begin{CJK*}{UTF8}{gbsn}祝伟\end{CJK*})}
\affiliation{Department of Astronomy, Tsinghua University, Beijing 10084, China}

\author[0000-0003-3015-6455]{Mu-Tian Wang (\begin{CJK*}{UTF8}{gbsn}王牧天\end{CJK*})}
\affiliation{Institute for Astronomy, University of Hawai‘i, 2680 Woodlawn Drive, Honolulu, HI 96822, USA}
\affiliation{School of Astronomy and Space Science, Nanjing University, Nanjing 210023, China.}
\affiliation{Key Laboratory of Modern Astronomy and Astrophysics, Ministry of Education, Nanjing, 210023, People’s Republic of China}

\author[0000-0003-3868-3663]{Max Goldberg}
\affiliation{Laboratoire Lagrange, UMR7293, Universit\'e C\^ote d'Azur, CNRS, Observatoire de la 
C\^ote d'Azur, Bouldervard de l'Observatoire, 06304, Nice Cedex 4, France}

\author[0000-0001-9985-0643]{Caleb Lammers}
\affiliation{Department of Astrophysical Sciences, Princeton University, 4 Ivy Lane, Princeton, NJ 08544, USA}

\author[0000-0003-1298-9699]{Kento Masuda \begin{CJK*}{UTF8}{min}(増田賢人)\end{CJK*}}
\affiliation{Department of Earth and Space Science, Graduate School of Science, Osaka University, 1-1 machikaneyama, Toyonaka, Osaka 560-0043, Japan}

\begin{abstract}
\noindent Recent observations suggest that the incidence of near-resonant planets declines as planetary systems age, making young planetary systems key signposts of early dynamical evolution. Here we investigate the dynamical states of three of the youngest multi-transiting planetary systems: AU Mic (3-planet, $\sim$20-Myr-old), V1298 Tau (4-planet, $\sim$23-Myr-old), and TOI-2076 (4-planet, $\sim$200-Myr-old). We find that most planet pairs in these systems lie near resonance with circulating rather than librating resonant angles. As a result, they are more susceptible to dynamical chaos than systems that are either securely locked in resonance or far removed from it. Even modest eccentricities of 0.04 to 0.08 may drive them to instability on timescales of tens to hundreds of Myr. Moreover, the observed orbital architectures are vulnerable to eccentricity excitation through mechanisms such as divergent resonance crossing triggered by planetesimal scattering. The observed near-resonant state may represent a transitional phase between a librating resonant chains and a mature non-resonant planetary system. Finally, we briefly discuss mechanisms that could give rise to the observed near-resonant configurations, including overstable libration, disk turbulence, and receding disk inner edge.
\end{abstract}


\section{Introduction} \label{sec:intro}

Disk migration naturally arises from gravitational interactions between forming planets and their natal protoplanetary disks \citep{Goldreich1979_excitation, Lin1986_Tidal, Kley_2012}. For low-mass planets, Type I migration (non-gap opening) generally gives rise to a drift towards the central star \citep{McNeil2005_Effects, Terquem2007_Migration}. As the innermost planet stalls near the disk inner edge \citep{Masset_2006,Wong2024_Resonant}, outer planets migrating inward can catch up, leading to convergent encounters in which the period ratio between neighboring planets decreases. Such convergent encounters frequently capture neighboring planets into mean-motion resonances \citep[e.g.,][MMRs]{Borderies1984_Simple, Lee2002}, locking their period ratios near integer commensurabilities (such as 2:1, 3:2). Resonant capture is more effective: 1) when planets begin on low-eccentricity orbits; 2) when migration proceeds adiabatically, i.e., the migration timescale is much longer than the resonant interaction timescale \citep{Borderies1984_Simple,Batygin2013_Analytical}. These conditions are often met in planet population synthesis models that incorporate Type I migration, thus these simulations often predict an abundance of mean-motion resonances among close-in planetary systems \citep[e.g., ][]{IdaLin2004a, IdaLin2004b, Mordasini2009a, Mordasini2009b, Ormel2017_Formation, Ogihara2018_Formation, Wong2024_Resonant, Keller2025_Higher-Order}.

In contrast, \textit{Kepler} planets exhibit a broad distribution of period ratios between neighboring planets, with only a small fraction found near first-order MMRs \citep[$\sim$15\%,][]{Fabrycky2014_Architecture, Winn_Fabrycky, Zhu2021_Exoplanet, Huang_Ormel, Dai2024_Prevalence}. To resolve the discrepancy with observations, the so-called ``Breaking-the-chains'' model was proposed \citep[e.g., ][]{Izidoro2017_Breaking,Ogihara2018_Formation,Pichierri2020_onset, Goldberg2022_Criterion,Li2024_Resonant}. In this scenario, orbital instabilities develop after the gaseous disk disperses, and may eventually break the initial resonances and reconfigure the orbital architecture. A similar evolution pathway has been proposed for the Solar System both for the gas giants \citep{Nesvorny2018,Liu2022} and more recently for the terrestrial planets \citep{Huang2025}. 

The dynamical evolution of the planetary systems, particularly the fraction of systems near MMRs, is therefore a crucial test of the ``breaking-the-chains'' scenario. \citet{Hamer2024_Kepler-discovered} found that \textit{Kepler} systems with near-resonant pairs, particularly those close to second-order MMRs, exhibit systematically lower velocity dispersions than field stars, suggesting younger stellar ages. Using direct age constraints from cluster membership and gyrochronology, \citet{Dai2024_Prevalence} found that near-resonant systems are nearly universal among young ($<$100-Myr-old) planetary systems, accounting for $\sim$80\% of the current sample. Moreover, the fraction of near-resonance planets declines with age, dropping to about 30\% for adolescent systems (0.1-1 Gyr) and further down to 15\% for mature systems \citep[$>$1 Gyr, see also][]{Fabrycky2014_Architecture,Huang2023_When}. These findings are qualitatively consistent with a scenario in which most planetary systems emerge from the disk phase as resonant chains, but subsequently evolve into near-resonant or non-resonant configurations through some dynamical processes.

While much of the discussion around MMR has focused on orbital period ratios, the defining characteristic of an MMR is the libration of a resonant angle within a separatrix, rather than proximity to an integer period ratio \citep[e.g., ][]{Batygin2015_capture}. A resonant angle is the generalized coordinate for the resonant part of a multi-planet Hamiltonian. For a 2-body resonance (2BR), it takes the form:
\begin{equation}\label{eqn:2-body}
    \phi_{\rm 12} =p\lambda_{2}-(p-q)\lambda_{1}-q\varpi_{1,2}.
\end{equation}
where $p$ and $q$ are positive integers and subscripts 1 and 2 respresents the inner and outer planet, respectively. The mean longitude $\lambda$ is the sum of the mean anomaly $M$, the longitude of the ascending node $\Omega$, and the argument of pericenter $\omega$. The angle $\varpi$ is the sum of $\Omega$ and $\omega$. A mixed pericenter angle $\varpi_{1,2}$ is used here, which is a weighted vector sum of $\varpi_{1}$ and $\varpi_{2}$, crucially it reduces the 2BR to one degree of freedom \citep[e.g., ][]{Sessin, Henrard1986, Wisdom1986, Batygin2013_Analytical, Hadden2019_Integrable}.  A necessary condition for true resonance is the libration of the resonant angle, i.e., finite-amplitude oscillation. Systems with near-integer period ratios but circulating angles (resonant angle goes through 0 to 2$\pi$) are formally near-resonant but not in resonance.

MMRs can either stabilize or destabilize planetary system, depending primarily on the dynamical state: whether the resonant angle librates or circulates. A classic example of stabilization is the 3:2 MMR between Neptune and Pluto \citep{Malhotra}. The libration of the resonant angle protects Neptune and Pluto from having close encounter despite Pluto’s eccentric, Neptune-crossing orbit. By contrast, asteroids in the Kirkwood gaps lack such librational protection; the combined influence of resonant and secular effects has cleared most asteroids near various MMRs \citep{Kirkwood_1867,Wisdom1983} \footnote{The 2:1 Kirkwood Gap is not completely depleted. The Zhongguo group inside this gap remains in libration, has a dynamical lifetime comparable to 4 Gyr\citep{Roig2002_zhongguo}.}. Similarly, mature resonant chains like TRAPPIST-1, which have remained stable over several Gyr, are also likely protected by librating MMRs \citep{Tamayo2017}.

In this work, we focus on characterizing the dynamical states and stability of the youngest confirmed multi-transiting systems, particularly those with transit timing variation (TTV) measurements that can directly probe resonant dynamics, i.e., AU Mic, V1298 Tau, and TOI-2076. 
In Section~\ref{sec:sys-select}, we summarize the key properties of the planetary systems analyzed in this study. Section~\ref{sec:near-res} and presents evidence that these systems are near, but not locked in, MMRs. Section~\ref{sec:not-in-3br} shows that these systems are not protected by three body resonance (3BR), either. 
Section~\ref{sec:stability} evaluates the long-term stability of the systems through N-body integrations. Finally, Section~\ref{sec:c_d} presents our conclusions and a discussion of the main results.

\section{System Selection}
\label{sec:sys-select}

We selected the youngest confirmed planetary systems with at least two pairs of planets close to low-order MMRs. Such configurations are unlikely to occur by chance, as population-level studies \citep[e.g., ][]{Huang2023_When, Dai2024_Prevalence} show that only $\sim$15\% of neighboring planet pairs lie close to first-order MMRs. The probability of two or more near-resonant pairs appearing coincidentally in a single system is low. We excluded two-planet systems due to the possibility of coincidental resonant alignment and the potential for incomplete architectures, as any undetected planets would significantly alter their dynamical stability. This left us with four planetary systems: AU Mic (3-planet, $\sim$20-Myr-old), V1298 Tau (4-planet, $\sim$23-Myr-old), TOI-2076 (4-planet, $\sim$200-Myr-old), and TOI-1136 (6-planet, $\sim$700-Myr-old).  TOI-1136 is excluded from further discussion as its dynamical state has already been well constrained in \citep{Dai2022_TOI-1136}. In short, the system appears to be in a librating state, with the possible exception of the second-order 7:5 resonance between planets f and g.

Crucially, detailed transit timing variation (TTV) analyses \citep{Agol2005} have been performed for all of our selected systems, yielding mass and eccentricity constraints that are essential for assessing their dynamical states. We summarize the key properties for AU Mic, V1298 Tau, and TOI-2076 below.

\begin{itemize}
    \item AU Mic is a pre-main-sequence (PMS) star that hosts two transiting planets \citep{Plavchan2020_planet} and a third non-transiting planet identified through TTVs \citep{Wittrock}. The host star is a member of the $\sim$20-Myr-old $\beta$ Pictoris moving group and is also known to harbor a debris disk \citep{Kalas2004_AUMicDisk}. Planetary masses and eccentricities in this system were determined using both TTV and Radial Velocity (RV) measurements \citep{Cale2021_Diving}. Here, we adopt the TTV posterior distribution from \citet{Wittrock}. \responseB{The masses of planets b and c are derived from the RV amplitudes in Table 18 of \citet{Wittrock}, assuming near-circular orbits, with their uncertainties conservatively set to approximately two-thirds of the best-fit values.}
    \item V1298 Tau is a PMS star in the Taurus-Auriga association ($\sim$23 Myr old). This system hosts four transiting planets \citep{David2019_Four}. For this study, we adopt the updated transit ephemerides, planetary masses, and eccentricities derived from a forthcoming TTV analysis by Livingston et al. (submitted). While RV mass estimates have previously been reported \citep{SuarezMascareno2021_Rapid}, their reliability has been challenged by \citet{Blunt2023_Overfitting} due to concerns over model overfitting. TTV analyses suggest that the planets in this system have low masses in the range of 5-15 $M_\oplus$, placing them in the sub-Neptune regime. This is consistent with the mass constraints from atmospheric modeling for the largest planet V1298 Tau b  \citep[12-15 $M_\oplus$, ][]{Barat2024_metal-poor,Barat2025}. We use the TTV masses from Livingston et al. (submitted) in our study.

    \item TOI-2076 is a PMS star hosting four transiting planets, with three sub-Neptunes b, c, and d \citep{Hedges2021_TOI-2076, Osborn2022_Uncovering}, and an inner super-Earth e \citep{Barber2025_TESS}. TOI-2076 comoves with TOI-1807 and other nearby stars; ensemble analyses place its age at $\sim$200-Myr-old \citep{Barber2025_TESS}. We adopt the planetary masses and eccentricities from a joint TTV-RV photodynamical model by Wang et al. (submitted).
\end{itemize}

As will be evident in the following sections, the key observables governing resonant dynamics are the orbital periods, planetary masses, and orbital eccentricities. Among these, orbital periods are the most precisely determined, with typical fractional uncertainties of less than 10 parts per million (ppm) \citep{Lissauer2024}. In contrast, planetary masses and eccentricities are less well-constrained and are subject to a well-known degeneracy in TTV analysis \citep{Lithwick2012_EXTRACTING}. Therefore, in the dynamical simulations of these systems, we explore a broader region in the mass and eccentricity space, while adopting the TTV posteriors as our baseline.

\begin{figure*}[ht!]
\centering
\includegraphics[width=1.0\textwidth]{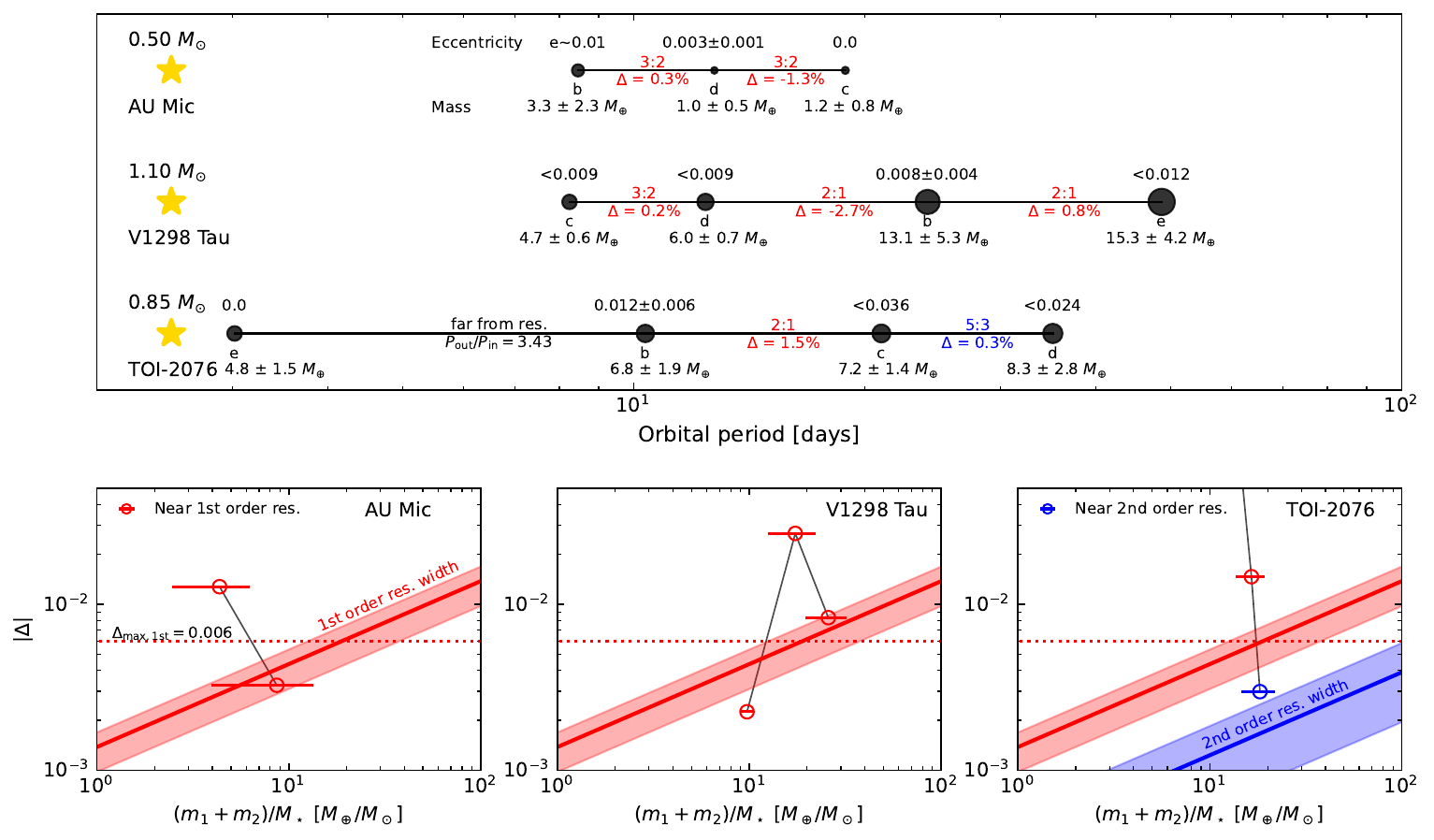}
\caption{
    {\bf Upper panel}: Orbital architecture of the young multi-planet systems studied here. The eccentricity and mass constraint from TTV modeling is shown above and below each planet, respectively \citep[Livingston et al. submitted, Wang et al. submitted,][]{Wittrock}. The size of each planet in the upper panel is proportional to its mass. {\bf Lower panels}: Pairwise period ratio deviation from exact commensurability ($|\Delta|$, Eqn \ref{eq:delta}) versus the planet-to-star mass ratio ($\mu = (m_1+m_2)/M_\star$).  Red and blue colors correspond to first-order ($q=1$) and second-order ($q=2$) MMRs, respectively. \responseB{The solid lines indicate the theoretical resonance width, $\Delta_{\max}$, calculated using Equation~(\ref{eq:res_width}) for a combined eccentricity of $|\tilde{e}| = 0.1$. The corresponding shaded regions show how this width varies as $|\tilde{e}|$ ranges from 0.05 to 0.15 for reference.} The horizontal red dotted line at $|\Delta| = 0.006$ represents the empirical libration threshold from \citet{Goldberg2023_Dynamics}. Each planetary system contains at least one pair whose $|\Delta|$ is outside the libration width, and is thus suggestive of a circulating state. Full TTV analysis also confirms the circulating states of these pairs with the possible exception of the innermost pair of V1298 Tau.
\label{fig:not-in-2br}
}
\end{figure*}

\section{Proximity to Pairwise 2-body Mean-Motion Resonances}
\label{sec:near-res}

We independently analyzed the TTV posterior distribution of AU Mic \citep{Wittrock},  V1298 Tau (Livingston et al. submitted) and TOI-2076 (Wang et al. submitted). Our analysis below shows that most planet pairs in these systems lie near resonance with circulating resonant angles, with the possible exception of the innermost pair in V1298 Tau, whose resonant state remains uncertain.

The proximity to a $p:(p-q)$ MMR is often quantified using the fractional deviation from exact commensurability, $\Delta$, defined as:
\begin{equation}
\Delta \equiv \frac{P_{\rm 2}/P_{\rm 1}}{p/(p-q)} - 1,
\label{eq:delta}
\end{equation}
where $P_{\rm 1}$ and $P_{\rm 2}$ are the orbital periods of the inner and outer planets, respectively.

Systems are typically considered near-resonant if $|\Delta|$ falls below 1-3\% \citep[e.g., ][]{Lithwick_repulsion, Batygin2013_Analytical, Choksi2020_Sub-Neptune}. Indeed, statistical studies of confirmed exoplanets reveal an excess of planet pairs with $|\Delta| \sim 1$-3\% just wide of first-order resonances \citep{Fabrycky2014_Architecture, Huang2023_When, Dai2024_Prevalence}. However, systems confirmed to be in true resonance, with librating resonant angles, generally exhibit a much smaller $\Delta \lesssim 10^{-3}$ \citep[e.g., ][]{Mills2016_resonant,Dai2022_TOI-1136}. N-body simulations of resonant chain formation also yield similarly small $\Delta$ values \citep[e.g., ][]{Ormel2017_Formation, Ogihara2018_Formation, Wong2024_Resonant, Keller2025_Higher-Order}, and hydrodynamical simulations provide $|\Delta|$ on the level of several parts per thousand \citep[e.g., ][]{AtaieeKley21}. These findings suggest that true resonance, or the libration of the resonant angle, typically requires much closer proximity to exact commensurability than is present in near-resonant systems for sub-Neptune planets. This conclusion is further supported by the population-level TTV analysis of \citet{Goldberg2023_Dynamics}, who identified an empirical boundary between resonant and near-resonant systems at $|\Delta|_{\rm max, emp} \approx 0.6\%$. They further noted that most \textit{Kepler} TTV systems \citep[e.g.][]{Hadden2017_Kepler} are near-resonant with circulating resonant angles.

To be more quantitative, we directly compare the observed $\Delta$ with the theoretical resonance width $\Delta_{\rm max, ana}$. The resonance width is the region in period ratio space around the nominal resonance where libration is the preferred solution. When $|\Delta| > \Delta_{\rm max, ana}$, libration becomes highly unlikely as the separatrix associated with the librating resonance no longer exists. 

In the limit of compact systems (i.e., period ratios near unity), the eccentricity dependence can be captured by a single quantity known as the relative eccentricity, denoted $\boldsymbol{e}_{12}$ \citep[e.g., ][]{Hadden2019_Integrable, Tamayo2024_Unified}. This parameter is a linear combination of the eccentricity vectors of the two planets and serves as a key control variable for determining the resonance width:
\begin{equation}
\boldsymbol{e}_{12} = \boldsymbol{Z}_2 - \boldsymbol{Z}_1 = e_2 e^{i \varpi_2} - e_1 e^{i \varpi_1}.
\end{equation}
Following the formalism in \citet{Tamayo2024_Unified}, the relative eccentricity is further normalized by the orbit-crossing eccentricity $e_{\rm c}$:
\begin{align}
\tilde{e} &= |\boldsymbol{e}_{12}|/e_{\rm c}, \\
e_c & \equiv \frac{(P_{\rm 2}/P_{\rm 1})^{2/3} - 1}{(P_{\rm 2}/P_{\rm 1})^{2/3} + 1}.
\end{align}
The approximate resonance width is then given by:
\begin{equation}
\Delta_{\rm max, ana} \approx 3 A_q \sqrt{\mu |\tilde{e}|^q},
\label{eq:res_width}
\end{equation}
where $\mu \equiv (m_1 + m_2) /M_\star$ is the planet-to-star mass ratio, and $A_q$ is a coefficient that depends on the resonance order $q$, with specific values, i.e., 0.84 and 0.75 for first and second-order resonance, taken from \citet{Tamayo2024_Unified}, respectively. 

In the lower panels of Figure~\ref{fig:not-in-2br}, we compare the observed values of $\Delta$ to the theoretical resonance widths $\Delta_{\rm max, ana}$ for the young planetary systems in our sample. Note that the innermost pair of TOI-2076 e and b is far from resonance and not shown in this plot. The resonance width is calculated over a wide range of planet-to-star mass ratios, assuming a representative normalized eccentricity of $|\tilde{e}| = 0.1$. \responseB{This corresponds to typical individual eccentricities of $e \sim 0.02$ and 0.01 for the 2:1 and 3:2 MMRs, respectively. These values are consistent with the constraints derived from TTV analyses, as shown in the upper panel of Figure~\ref{fig:not-in-2br}. The shaded bands in Figure~\ref{fig:not-in-2br} show how the resonance width varies for $|\tilde{e}|$ between 0.05 and 0.15, with individual eccentricities scaling accordingly.}

Figure~\ref{fig:not-in-2br} shows that each planetary system contains at least one pair whose observed $|\Delta|$ significantly exceeds both the calculated resonance width $\Delta_{\rm max, ana}$ and the empirical libration threshold of $|\Delta|_{\rm max, emp} \approx 0.6\%$ proposed by \citet{Goldberg2023_Dynamics}. In AU Mic and V1298 Tau, there is one planet pair that has a large negative $\Delta$ value, placing it even farther from the nominal resonance. In TOI-2076, the planet pair near 2:1 first-order resonance shows a $|\Delta|$ well above the expected resonance width, indicating that libration is unlikely. While the TOI-2076 pair near the second-order 5:3 resonance exhibits a smaller $|\Delta|=0.3\%$, it still exceeds the width estimated for this weaker resonance (indicated by the blue dashed line), reinforcing the conclusion that the system is not in resonance.

To quantify the systematic uncertainty introduced by the pendulum approximation adopted in \citet{Tamayo2024_Unified}, we compared the resonance width calculated using Equation \ref{eq:res_width} with the more complete treatment of \citet{Hadden2019_Integrable}. For the low eccentricities relevant here, the systematic error in resonance width is of order $\sim$0.1\%. This is comparable to the observational uncertainties in planet masses and eccentricities inferred from TTV analyses and therefore does not affect our qualitative conclusions. We also considered that the libration center could shift to a value slightly larger than commensurability in the low-eccentricity limit, but this possibility of forced libration is inconsistent with the observed negative $\Delta$ values and the absence of 3BR.

The dominant periodicity observed in the TTV signals offers further insight into the dynamical state of a planetary system. In near-resonant systems, the characteristic TTV signal is governed by the so-called super-period $P_s= 1/|p/P_{\rm 2}-(p-q)/P_{\rm 1}|$, which scales as $P_{\rm orb}/\Delta$. In contrast, for systems that are truly in resonance, the dominant periodicity corresponds to the libration period $P_l \approx P_{\rm orb} (\frac{m_1+m_2}{M_\star})^{-2/3}$ \citep[e.g., ][]{Lithwick2012_EXTRACTING, Nesvorny2016_DYNAMICS}. Except for the innermost pair in V1298 Tau, the observed TTV signals for all three systems generally align with the expected super-periods \citep[Wang et al. submitted; Livingston et al. submitted;][]{Wittrock}. Even without a full N-body integrations, both the resonance width and the TTV periodicities indicate that the neighboring planet pairs in AU Mic, V1298 Tau, and TOI-2076 are not locked in 2BR resonances.
 
\section{Proximity to 3-Body, Laplace-like Resonances}
\label{sec:not-in-3br}

\responseB{Neighboring triplets of planets can be engaged in 3-body Laplace-like resonances (3BRs), with the most famous example being the resonance among Jupiter's Galilean moons \citep{Murray1999_Solar}. A 3BR can exist even if neighboring pairs are not locked in 2BRs; such a configuration has been proposed for the Kepler-221 system \citep{Goldberg2021_Tidal,Yi2025_Dynamical}.} \citet{Agol2021} also demonstrated that 3BR resonant angles are all librating for neighboring triplets in TRAPPIST-1, while 2BR resonant angles may be circulating. The libration of a 3BR resonant angle prevents triple conjunctions and the associated chaotic interactions, thereby promoting long-term dynamical stability.

The 3BR angle for a triplet of planets can be defined as
\begin{equation}
\phi_{\rm 3BR} = k_1 \lambda_1 - k_2 \lambda_2 + k_3 \lambda_3
\label{eq:phi_3br}
\end{equation}
where $\lambda_i$ are the mean longitudes and $k_i$ are integer coefficients. For zeroth-order 3BR, $k_1 - k_2 + k_3 = 0$ to satisfy the d'Alembert relation \citep{Murray1999_Solar}. We ignored higher order 3BR. If a triplet is locked in a zeroth-order 3BR, the corresponding resonant angle librates. A necessary, though not sufficient, condition for libration is that the time derivative of the resonant angle is small:
\begin{equation}
B = \dot{\phi}_{\rm 3BR} = k_1 n_1 - k_2 n_2 + k_3 n_3
\label{eq:B_3br}
\end{equation}
where $n_i = 2\pi/P_i$ is the mean motion of planet $i$. To compare the degree of proximity to a 3BR across systems, a normalized quantity $B_{\rm norm} = |B|/\langle n \rangle$ is used here, where $\langle n \rangle$ is the average of the mean motions for the triplet \citep{Goldberg2021_Tidal}. 

\begin{figure*}[ht!]
\centering
\includegraphics[width=0.9\textwidth]{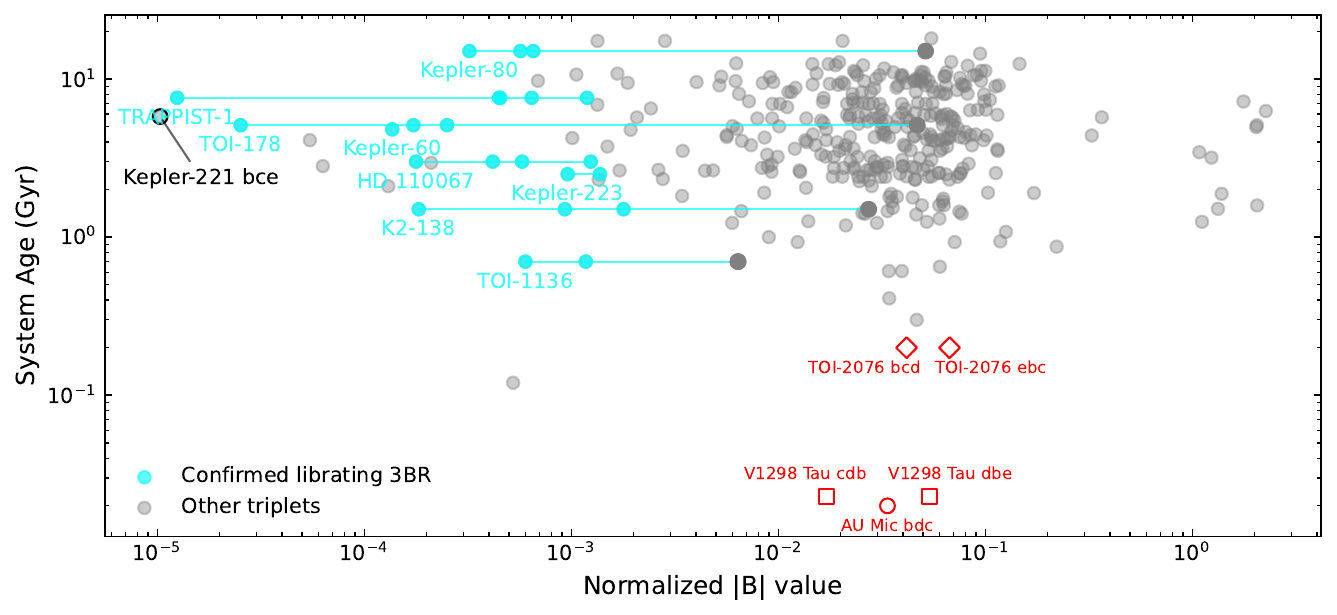}
\caption{
    Proximity to 3-body Laplace-like resonance $|B_{\rm norm}| = |k_1 n_1 - k_2 n_2 + k_3 n_3| / \langle n \rangle$ for neighboring triplets of planets versus system age. Triplets with reported librating 3BR angles, either from TTV or from dynamical analysis, are shown in cyan (Kepler-80 \citep{MacDonald2016_DYNAMICAL}, TRAPPIST-1 \citep{Gillon2017_Seven}, TOI-178 \citep{Leleu2021_Six}, Kepler-60 \citep{Gozdziewski2016_Laplace}, HD 110067 \citep{Luque2023_resonant, Lammers2024_Six-planet}, Kepler-223 \citep{Mills2016_resonant}, K2-138 \citep{Christiansen2018, MacDonald2022_K2_138}, TOI-1136 \citep{Dai2022_TOI-1136}). All other triplets without a confirmed librating 3BR angle are shown in grey. The non-adjacent Kepler-221 b-c-e triplet, which is suspected to be librating, is also included \citep{Goldberg2021_Tidal, Yi2025_Dynamical}. The plot reveals two distinct populations: the librating triplets (cyan) have $|B_{\rm norm}|$ values approximately two orders of magnitude smaller than the non-librating general population (grey). The young planetary systems (AU Mic, V1298 Tau, TOI-2076; shown in red) have triplets whose $|B_{\rm norm}|$ are so large that the resonant angle is most likely circulating. \responseB{We omit the non-adjacent triplets for V1298 Tau and TOI-2076 for clarity and note that they are all far from 3BR with $|B_{\rm norm}| > 0.02$.}
\label{fig:not-in-3br}}
\end{figure*}

For each adjacent triplet in the sample, we searched for the minimum $|B|$ values over many possible zeroth-order three-body resonance configurations with integer coefficients $k_i$ ranging from 0 to 10. The corresponding minimum normalized $|B_{\rm norm}|$, was then computed and used as a metric for comparison across systems. To assess the likelihood of a 3BR in AU Mic, V1298 Tau, and TOI-2076, we compare their minimum $|B_{\rm norm}|$ values against triplets with confirmed librating 3BRs (see Figure~\ref{fig:not-in-3br} for the full list) as well as the general population of confirmed planetary triplets. We compiled a comparison sample of transiting multi-planet systems from the NASA Exoplanet Archive\footnote{https://exoplanetarchive.ipac.caltech.edu/} as of May 2025, selecting all adjacent planetary triplets. We also included the non-adjacent triplet Kepler-221 b-c-e, which exhibits an exceptionally small $B$ value indicative of a pure 3BR \citep{Goldberg2021_Tidal, Yi2025_Dynamical}. 

We plot the normalized $B$ value against the stellar age in Figure~\ref{fig:not-in-3br}. The age of the system is assigned either from \citet{Berger_age} or from \citet{Dai2024_Prevalence} and references therein. For systems without a reported age, we assign a random value between 1 and 10 Gyr for the purpose of illustration. The ages of two mature resonant chain systems, HD 110067 and TOI-178, are shifted slightly to avoid clutter. Planetary triplets from systems with reported librating three-body resonances are highlighted in cyan. The relevant references are provided in the figure caption. Other triplets that are either non-resonant or lack detailed dynamical characterization are shown in grey. The triplets from the young systems (AU Mic, V1298 Tau, and TOI-2076) are indicated by red hollow markers.

Figure~\ref{fig:not-in-3br} reveals a clear separation between two populations: systems with confirmed 3BR libration exhibit $|B_{\rm norm}|$ values that are typically orders of magnitude lower than those of the general population. AU Mic, V1298 Tau, and TOI-2076 have $|B_{\rm norm}|$ values similar to non-librating general population. This indicates that the 3BR in these three young systems are unlikely to be librating.

\begin{figure*}[ht!]
\centering
\includegraphics[width=1.0\textwidth]{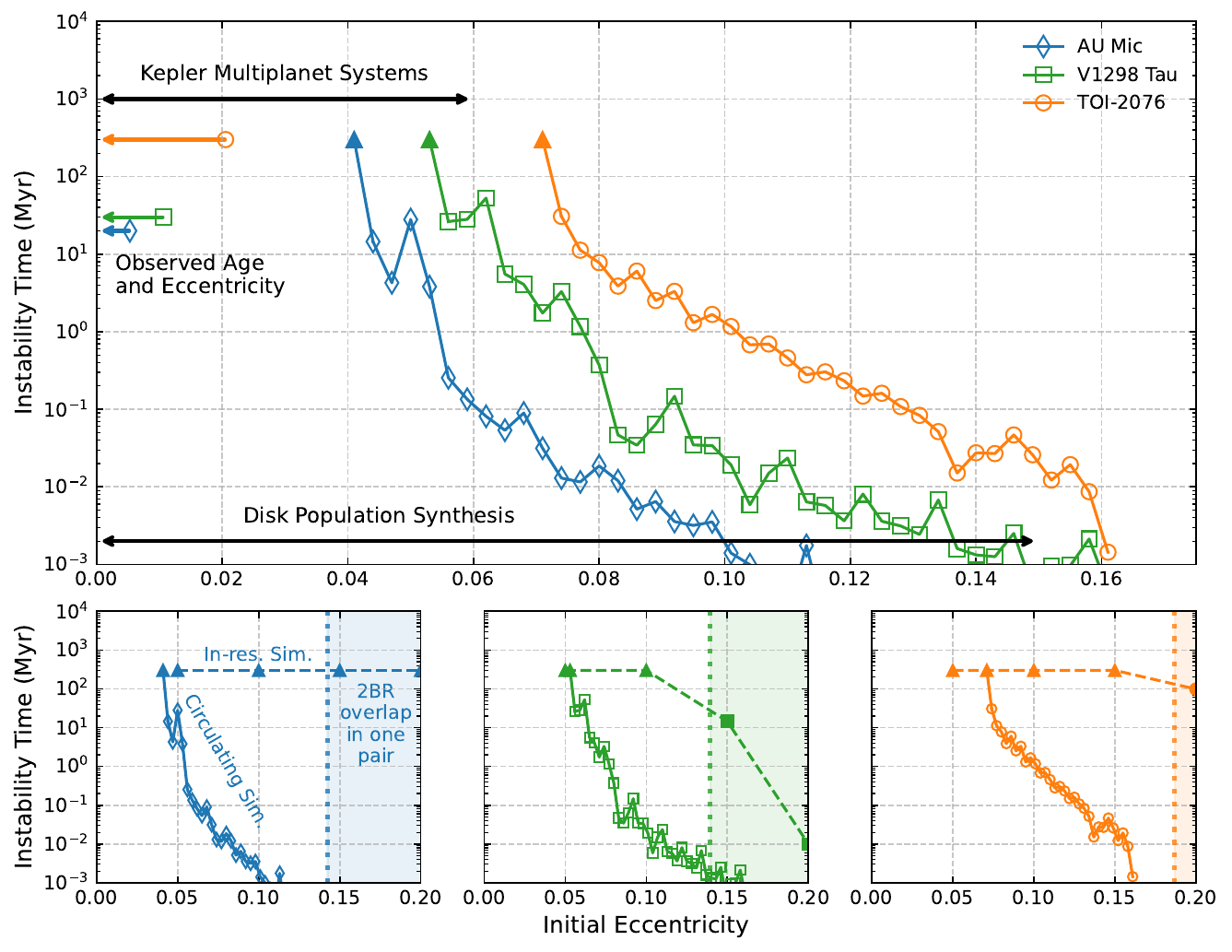}
\caption{\textbf{Top panel:} Instability timescale as a function of initial eccentricity. Filled triangles indicate that the instability timescale exceeds the simulation duration of 300 Myr. Hollow points with left arrows denote observed eccentricity upper limits measured via TTVs and system age for each system. The upper horizontal double arrow shows the eccentricity range for observed planet systems \citep{Hadden2017_Kepler, Xie2016_Exoplanet, VanEylen_multi2015, Gilbert_Petigura}.
The lower horizontal double arrow shows the eccentricity range from N-body migration simulations \citep{Keller2025_Higher-Order, Izidoro2017_Breaking}. \textbf{Bottom panels:} Comparison between instability timescales for the near-resonant \responseB{(hollow symbols/lines)} and in-resonance configurations (dashed lines). Near-resonant systems are much more susceptible to orbital instability. Shaded regions indicate the critical eccentricity beyond which neighboring 2BRs begin to overlap \citep{Deck2013_overlap, Hadden2018_Criterion}. Even in-resonant system become unstable beyond this point.
\label{fig:e-inst}
}
\end{figure*}

\section{The Vulnerable Young Planet Systems}
\label{sec:stability}

\responseB{To assess the orbital stability of each system, we performed a suite of N-body integrations using the symplectic Wisdom-Holman integrator \texttt{WHFast} \citep{Wisdom1991_WHFast, Rein2015_WHFast} within the \texttt{REBOUND} package \citep{Rein2012_Rebound}, neglecting tidal forces and general relativistic effects.} We varied the planetary masses and orbital eccentricities over a wider range than the TTV posterior distribution. \responseB{In these integrations, all planets in a system were assigned a common initial eccentricity, or their best-fit masses were scaled by a common factor. To isolate the impact of each parameter, we varied either the eccentricities or the masses while holding the other parameter fixed at its TTV-derived best-fit value.} The initial values of $\varpi_i$ and $\lambda_i$ were randomly drawn to reflect the circulating, non-resonant nature of the observed configurations. At each point in the parameter grid, we ran 30 independent integrations and report the median instability timescale.

Our first result is that if AU Mic, V1298 Tau, and TOI-2076 can maintain the low orbital eccentricities as inferred from TTV analyses ($e \lesssim 0.02$, isolated markers in Figure~\ref{fig:e-inst}), the systems remain stable for timescales exceeding 300 Myr, longer than the approximate age of TOI-2076 and the duration of our integrations. We consider a system to have gone unstable if any two planets experience orbital crossing or a close encounter within a mutual Hill radius.  These two instability criteria yield comparable instability timescales across the ensemble \citep[e.g., ][]{PuWU, Lammers2024_Instability}. This result confirms that, in their current configurations, the systems are not at risk of imminent disruption.

However, if AU Mic, V1298 Tau, and TOI-2076 were to acquire moderate eccentricities ($\gtrsim 0.04$), they could go unstable within the next $\sim$100 Myr. The top panel of Figure \ref{fig:e-inst} reveals a steep inverse relationship between instability timescale and eccentricity. \citet{Yee2021} analyzed the orbital stability of \textit{Kepler} planets with TTVs, most of which are also near-resonant. They found the unstable probability of these TTV planets steeply increase with eccentricity. Given that the observed decline in near-resonant system populations occurs on timescales of $\sim$100 Myr \citep{Dai2024_Prevalence,Schmidt}, this steep dependence of stability on eccentricity is particularly relevant for the evolution of initially resonant systems.

We briefly place the observed low orbital eccentricities ($\lesssim$0.02) of AU Mic, V1298 Tau, and TOI-2076 into a broader context. In N-body disk migration simulations, equilibrium eccentricities arise from a balance between resonant excitation and damping by the gaseous disk. For sub-Neptunes and a wide range of damping efficiencies, values of $e \sim 0.01-$0.10 have been reported \citep[e.g.,][]{Keller2025_Higher-Order}. In `breaking-the-chains’ scenarios, eccentricities can be excited to $e \sim 0.1$ following close encounters or collisions \citep{Izidoro2017_Breaking}. Finally, the eccentricity distribution of mature, non-resonant {\it Kepler} planets have been reported to be around 0.03 to 0.06 \citep[e.g., ][]{VanEylen_multi2015, Xie2016_Exoplanet, Hadden2017_Kepler, Gilbert_Petigura}. It is unclear why AU Mic, V1298 Tau, and TOI-2076 currently have low eccentricities. Nonetheless, it is plausible that they may undergo a phase of eccentricity excitation and hence enhanced instability in the future: a possibility we examine further in Section ~\ref{subsec:e-excite}.

As expected, we found that near-resonant systems are much more susceptible to orbital instability than their in-resonance counterparts. The second row of Figure~\ref{fig:e-inst} compares the orbital stability of near-resonant and in-resonance configurations with otherwise identical system parameters. \responseB{To initialize these resonant systems, we placed each adjacent planet pair near the libration center by anti-aligning their $\varpi_i$ and adjusting the initial mean longitudes ($\lambda_i$) such that the resonant angle $\psi \approx \pi$. We then introduced small perturbations to the mean longitudes, drawing them from a normal distribution with a standard deviation of 10$^\circ$ ($\Delta \lambda_i \sim \mathcal{N}(0, 10^\circ)$).} The in-resonance systems remain stable over the full duration of our simulations, except when eccentricities grow large enough for two-body resonance overlap to occur \citep[$e$ of order 0.15, ][]{Deck2013_overlap, Hadden2018_Criterion}.

Furthermore, systems positioned far from either 2BR or 3BR resonance exhibit enhanced stability compared to the near-resonant cases. For instance, the instability timescale for AU Mic is around 10 Myr with eccentricity of 0.05. If one increase the period ratio of the inner planet pair by just 1\%, which would move it sufficiently far from resonance, the instability timescale is extended to more than 300 Myr.

\begin{figure}[ht!]
\centering
\includegraphics[width=1.0\columnwidth]{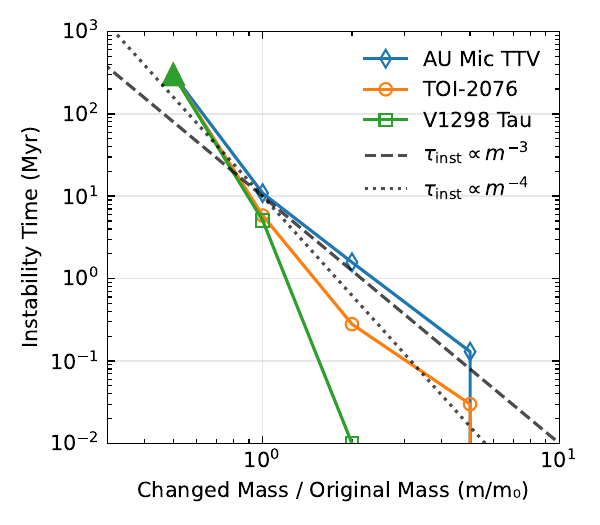}
\caption{
    The instability timescale versus planetary masses. The markers are the same as in Figure~\ref{fig:e-inst}. The simulations were initialized with an eccentricity where each system has an instability time of approximately 10 Myr with its measured masses. The dependence of the instability timescale on planetary mass is steep, likely between the 3rd or 4th power of the planetary mass.
\label{fig:m-inst}
}
\end{figure}

The orbital stability of near-resonant systems such as AU Mic, V1298 Tau, and TOI-2076 also shows a strong dependence on planetary masses, as illustrated in Figure~\ref{fig:m-inst}. Qualitatively, increasing planetary masses while holding other orbital parameters fixed reduces the interplanetary spacing in units of mutual Hill radii or $(M_p/M_\star)^{-1/4}$ as suggested by \citet{Petit2020_path, Lammers2024_Instability}, thereby driving the system toward instability \citep[e.g., ][]{PuWU}. Quantitatively, our simulations reveal that the instability timescale scales inversely with planetary mass to approximately the third or fourth power. \responseB{Consistent with predictions for 3-body resonance overlap in more compact systems \citep{Quillen2011, Petit2020_path, Lammers2024_Instability}, we find that the relationship between instability timescale and mass also extends to the less compact systems investigated in this study.}

This discrepancy arises because our systems bypass the typical bottleneck for instability driven by 3BR diffusion. Recent studies suggest that orbital instability of an initially non-resonant multi-planet system is a two-step process \citep{Petit2020_path, Lammers2024_Instability}: 1) the system experiences slow chaotic diffusion of semi-major axes driven by overlapping 3BRs, followed by 2) rapid onset of instability and orbit crossings once diffusion brings planets to a strong first-order 2BR. The bottleneck is usually the initial chaotic diffusion. The systems we studied here are already near strong 2BRs, bypassing the slow diffusion stage entirely.

\section{Summary and Discussion}
\label{sec:c_d}

\subsection{Potential Evolution Pathways}
\label{subsec:evo_path}
In the classical ‘breaking-the-chains’ model \citep[e.g.,][]{Izidoro2017_Breaking}, one might naively expect two types of orbital architectures: (1) young, librating resonant chains produced by convergent disk migration, and (2) mature, non-resonant systems that have already undergone resonant disruption and orbital instability. However, the three young systems studied here may represent some intermediate state, where planet pairs are near but not in resonance, and the orbital eccentricities are low. As a result, the system are mildly chaotic but not in any immediate danger of going unstable. We offer some explanations for how AU Mic, V1298 Tau, TOI-2076 arrived at the currently observed dynamical state.

Resonant capture may be temporary, with planets frequently escape MMR due to overstability. During convergent disk migration, planets can be captured into mean-motion resonance. However, when dissipative forces of the gaseous disk is properly accounted for, the resonant fixed point may become unstable in the sense that small deviations from the fixed point grow exponentially. This so-called `overstable libration' grows in amplitude until the planets leave the resonance \citep{Goldreich2014, Deck_Batygin, Xu0217, Lin2025_capture}. Overstability is favored when the outer planet is more massive than the inner one, and is particularly effective near the 2:1 resonance. After escaping MMR through overstability, a planet pair is often left in the so-called ‘inner circulation’ regime, where the period ratio lies just short of an integer ratio ($\Delta<0$), and eccentricities are damped to nearly zero. The middle pair of V1298 Tau (planet d and b, see Figure \ref{fig:not-in-2br}) fits these descriptions almost perfectly: they are near 2:1 resonance; the outer planet is about twice more massive than the inner planet; the currently observed $\Delta=-2.7\%<0$; their eccentricities are small ($e\lesssim0.01$).

One caveat, however, is that the entire process of overstability and escape into inner circulation occurs on a timescale comparable to $\tau_e$ the eccentricity damping timescale. Eccentricity damping timescale is likely much shorter than the disk lifetime \citep[e.g.][]{Nelson2018}. To explain V1298 Tau via overstability, disk dispersal would therefore need to occur at a finely tuned moment: when the resonant-excited eccentricity have been sufficiently damped, but before the pair has migrated too far from the escaped 2:1 resonance. One possible solution is that if the $\tau_e$ varies with the disk properties. In the last vestige of the disk, $\tau_e$ may increase with the decreasing surface densities, and therefore alleviate the timing problem. We leave a thorough exploration of this scenario to a future study.

In traditional disk migration simulations, disks are assumed to be laminar. However, stochastic forcing from disk turbulence can also hinder resonant capture by broadening the parameter space required for overstable libration \citep[e.g., ][]{Soto, Chen2025_Capture}. It is also the case that spiral density waves excited by planets can interfere with each other, generating additional torques that inhibit resonance trapping \citep[e.g., ][Huang et al. in prep]{Yang2024_Provided}. This mechanism can produce planet pairs wide of resonance, i.e., with positive $\Delta$, but struggles to explain the negative $\Delta$ values observed in some of the selected systems. 

The expansion of the inner disk edge can modify the dynamical state of a resonant chain parked at the disk inner edge \citep{Masset_2006,Wong2024}. Planets inside of the disk inner edge primarily interact with the outer Lindblad resonances, and thus tend to migrate inward. Whereas, planets just outside the disk inner edge effectively feels an outward torque due to the dominating effect of co-rotation torque \citep{Masset_2006}. As the disk edge expands, a divergent migration between planets inside and outside the disk egdge takes place  \citep[e.g., ][]{Liu2017,Liu2022,Huang_Ormel,Pichierri2024,Hansen2024}. Such a divergence can disrupt the librating state of a resonant chain if the recession rate is sufficiently rapid. We encourage follow-up studies to work out the details in the context of AU Mic, V1298 Tau, and TOI-2076.

After the gas disk dissipates, additional dynamical processes can displace planets from resonance. Examples include planetesimal scattering \citep[e.g., ][]{Chatterjee_2015, Raymond2021, Wu2024_Repelling}, tidal dissipation \citep[e.g., ][]{Lithwick_repulsion, Batygin_repulsion, Lee2013, Millholland_obliquity}, or atmospheric mass loss \citep[e.g., ][]{Matsumoto2020_Breaking,WangLin,Hanf}. However, tidal evolution and atmospheric mass loss typically operate on long timescales of order $\sim$100 Myr. Given the youth of the systems studied here, especially 20-Myr-old AU Mic and V1298 Tau, these mechanisms may not have had enough time to operate. We defer a discussion of their potential influence on the long-term evolution of these systems to the next subsection.

\subsection{Mechanism for Eccentricity Excitation}
\label{subsec:e-excite}

The currently eccentricities in young planetary systems AU Mic, V1298 Tau, and TOI-2076 are sufficiently low to promote dynamical stability for at least 300 Myr. Nonetheless, several mechanisms can excite eccentricities over time, potentially driving these systems toward orbital instability and the eventual disruption of their near-resonant configuration.

When a pair of planets divergently cross a mean-motion resonance (i.e. when their period ratio increases over time), they receive a dynamical kick that excites both their eccentricities and their period ratio \citep[e.g., ][]{Murray1999_Solar, Wu2024_Repelling, Lin2024_Creating}. This mechanism is especially relevant for the near-resonant young systems studied here, particularly for pairs with negative $\Delta$ e.g. V1298 Tau db and AU Mic dc (Figure~\ref{fig:not-in-2br}). These pairs require only a small amount of divergent migration to encounter an MMR. The amount of eccentricity excitation during such a divergent encounter can be accurately predicted. This is because during a divergent encounter, the adiabatic invariant is broken, instead the system adopts the phase space area of the separatrix at bifurcation \citep{Murray1999_Solar}. The phase space area directly relates to the orbital eccentricity. Following \citet{Wu2024_Repelling}, a pair of planets divergently crossing a $p:(p-1)$ resonance obtain an eccentricity:

\begin{equation}
\begin{aligned}\label{eqn:div}
\Delta e_{12} &\approx 4\% \frac{[(p-1)/2]^{1/3}}{(p/3)^{2/3}} \left( \frac{M_1 + M_2}{16 M_\oplus}\right)^{1/3} \left( \frac{M_\star}{M_\odot}\right)^{-1/3},
\end{aligned}
\end{equation}
where the normalization assumes two 8 $M_\oplus$ planets orbiting a solar-mass star at the 3:2 resonance. Notably, the $\sim$0.04 eccentricity excited in this case is comparable to the level required to trigger orbital instability in our N-body simulations (Section~\ref{sec:stability} and Figure~\ref{fig:e-inst}).

Following disk dispersal, planetesimal scattering and atmospheric mass loss could both induce the required divergent migration \citep{Chatterjee_2015,Raymond2021,Wu2024_Repelling}. Young systems could retain debris disks, as exemplified by the edge-on disk of AU Mic \citep{Kalas2004_AUMicDisk}. Gravitational scattering of planetesimals drives systematic orbital migration, with the fractional change in period ratio proportional to the planetesimal-to-planet mass ratio $\epsilon$ \citep[i.e., $\Delta \left( {P_1}/{P_2}\right) \approx 1.1 \epsilon$,][]{Wu2024_Repelling}.
Given the proximity to resonance in these systems, planetesimals comprising merely 1-2\% of the planetary mass could induce sufficient migration to trigger resonance crossing and subsequent eccentricity excitation \citep[see also][]{Raymond2021}.

We simulated a toy model of divergent migration of AU Mic using \texttt{REBOUNDx} \citep{Tamayo2019_REBOUNDx}. We applied migration timescales of $\tau_{a} = -200$ Myr (inward) on planet d and $\tau_{a,c} = 100$ Myr (outward) on planet c, calibrated to achieve a $\sim$1\% period ratio change over 1 Myr. This corresponds to the effect of scattering by a planetesimal population with a mass of order 1\% of the planets. As the planet pair crosses the 3:2 resonance, the relative eccentricity reached $\sim$0.05, consistent with the values predicted by Equation \ref{eqn:div}.

Young planets are expected to undergo substantial atmospheric erosion through boil-off \citep{Owen_boiloff}, photoevaporation \citep{Owen,Lopez2014} and core-powered mass loss \citep{Ginzburg}. To reproduce the observed bimodal radius distribution of sub-Neptune planets, young planets such as those in AU Mic, V1298 Tau, and TOI-2076 are expected to eventually shed a few to tens of percentile of their masses in H/He \citep[e.g.][]{Owen_review}. If the mass loss is anisotropic, it can impart a net change in angular momentum to the planet. Although the efficiency of this process remains debated \citep[cf.][]{Hanf, WangLin}, it is plausible that mass loss could drive an increase in orbital eccentricity. Furthermore, if the mass-loss-induced orbital migration leads to a divergent resonance crossing, the eccentricity excitation described by Equation~\ref{eqn:div} would also apply.

A final possibility is that systems like AU Mic, V1298 Tau, and TOI-2076 also host longer-period giant planets with substantial angular momentum deficit \citep{Laskar2017_AMD-stability}. It is estimated $\sim30\%$ of compact, multi-planet systems have outer giant planets \citep[e.g.][]{Zhu2018}. Secular interactions, particularly secular chaos \citep[e.g.][]{Lithwick2011,Petrovich,Huang2025} may transfer angular momentum deficit from the outer companions to the inner planets and excite their orbital eccentricities. Since giant planets are much more massive, even a small eccentricity would produce a substantial angular momentum deficit that can then be equi-partitioned to the inner planets. At present, however, observational data are insufficient to confirm or rule out the presence of such outer companions in AU Mic, V1298 Tau, and TOI-2076.

\subsection{Summary}

In this paper, we investigated the dynamical state and long-term stability of the three youngest, multi-planet systems: AU Mic, V1298 Tau, and TOI-2076. We found these systems are most likely not locked in either two-body or three-body resonant chains despite their near-commensurate period ratios. Instead, most of the planet pairs in these systems (with the possible exception of the innermost pair of V1298 Tau) exhibit near-resonant configurations characterized by circulating resonant angles. None of the 3BR resonant angles appear to be librating in these systems.

This dynamical state has significant implications for their stability. AU Mic, V1298 Tau, and TOI-2076 can remain stable for more than 300 Myr thanks to their measured low eccentricities ($e \lesssim 0.02$). However, our simulations suggest mild eccentricity excitation to 0.04-0.08 could push these systems to orbital instability that unfolds in the next 10s to 100s of Myr. The strong sensitivity of instability timescales to both eccentricity and planet-to-star mass ratio suggests that near-resonant configurations represent an intermediate, metastable stage in planetary system evolution. By contrast, if these systems were in librating resonance, they would remain stable until $e \gtrsim 0.15$ where resonance overlap occurs.

We described a few potential mechanisms that could excite the orbital eccentricity: divergent encounter induced by planetesimal scattering or mass loss, or secular interaction with longer-period companions. The dynamical state (i.e., locked-in or near resonance) of young planets is critical for connecting young planetary systems with the mature ones, the latter appear to be near the edge of dynamical stability \citep[e.g., ][]{PuWU, Zhu2021_Exoplanet}. Recent observations further underscore the role of dynamical processes in shaping planetary architectures, showing that the fraction of near-resonant pairs declines systematically with system age \citep{Dai2024_Prevalence, Hamer2024_Kepler-discovered}.

Our conclusion here are necessarily limited by small-number statistics and potential observational biases. One important caveat is that TTV amplitudes scale as $1/|\Delta|$ \citep[e.g., ][]{HaddenLithwick2016}, planet pairs closer to resonance tend to exhibit larger TTV amplitudes and are therefore systematically harder to detect in periodic transit searches. This introduces an observational bias against identifying truly resonant planets. Therefore, expanding the sample size of well-characterized young planetary systems, as well as searching for additional planets in known young systems, will be essential for testing and refining our understanding of dynamical evolution.

\section*{acknowledgements}

\responseB{We thank the anonymous reviewer for comments and constructive suggestions.} We thank \responseB{Erik Petigura}, Sam Hadden, Yanqin Wu, Beibei Liu, Douglas N. C. Lin, Konstantin Batygin, Justin Wittrock, Peter Plavchan, Chris Ormel, Shuo Huang, Tian Yi, Andrew Mann, Yixian Chen, Andre Izidoro, Sean Raymond, Brad Hansen, and Allona Vazan for their helpful discussions. Work by ZH and WZ is supported by the National Natural Science Foundation of China (grant No. 12173021
and 12133005). 

\vspace{5mm}

\appendix

\section{Chaos Induced by Near Resonance}
\label{sec:chaotic}

\begin{figure*}[ht!]
\centering
\includegraphics[width=1.0\textwidth]{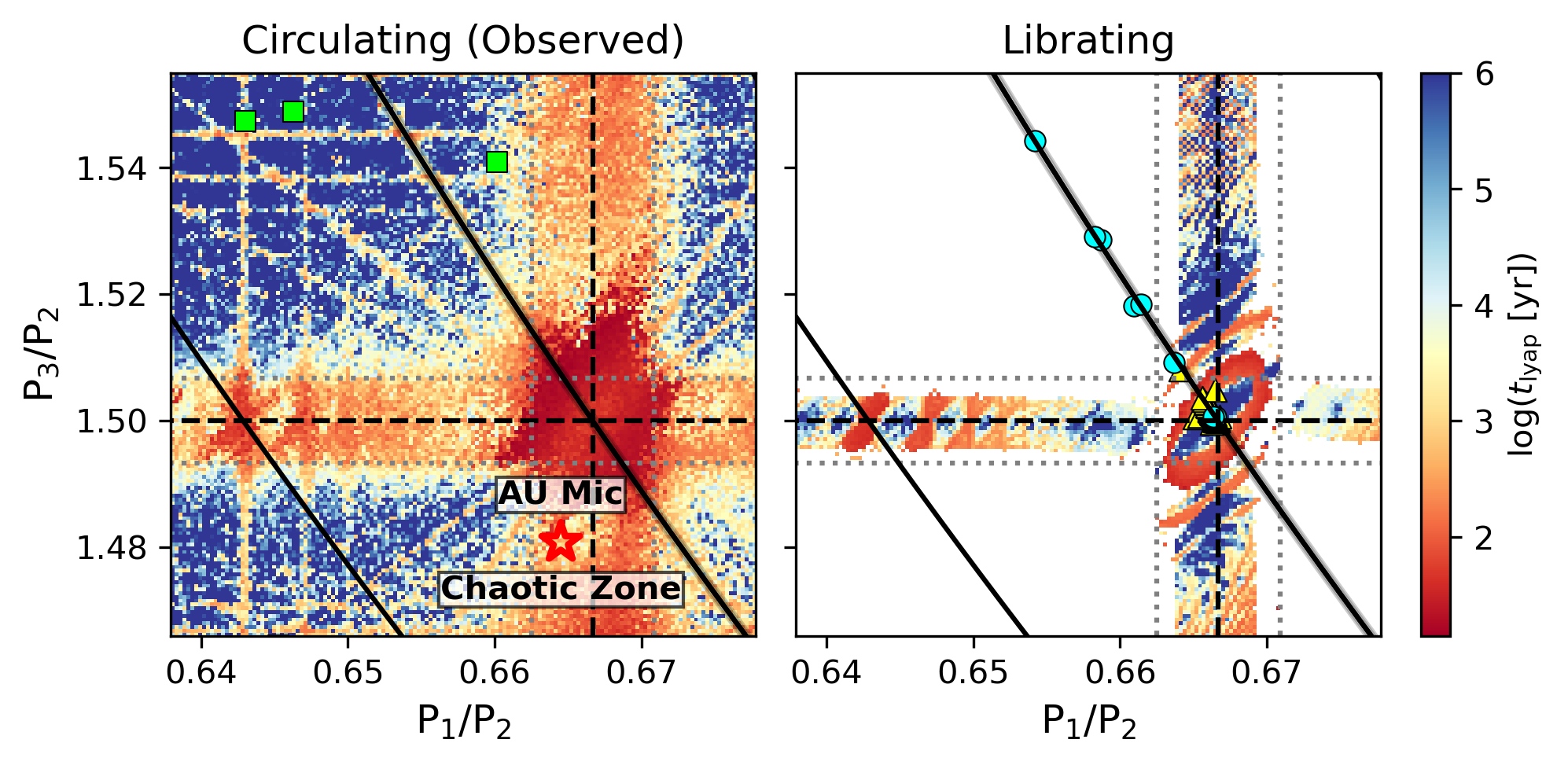}
\caption{
Lyapunov timescale maps for the AU Mic system with circulating (left) and librating (right) resonant angle at e=0.04. The color scale represents the logarithm of the Lyapunov timescale in years, with warmer colors indicating shorter timescales and more chaotical behavior. The red star marks the nominal position of AU Mic. Given that AU Mic likely exhibits circulating behavior near resonance, it is only shown in the left panel. Cyan circles denote confirmed librating 3BR systems (e.g. TRAPPIST-1), while green squares represent observed non-resonant multi-planet systems (corresponding to systems shown in Figure \ref{fig:not-in-3br}). Yellow triangles indicate resonant chains from N-body migration simulations \citep{Keller2025_Higher-Order}. Black solid lines mark the locations 3BR, with shaded regions showing estimated resonance widths \citep{Petit2020_path}. The clustering of cyan points near these lines suggests the dynamical protection offered by librating 3BRs. The dashed black line and dotted gray line indicate the nominal 2BR location and width computed by Equation \ref{eq:res_width}. Notably, the deep resonance region near the 2BR, which appears regular under libration, becomes chaotic under circulation. These maps reveal that the near-resonant configuration of AU Mic places it within a region of substantial dynamical chaos.
\label{fig:lyap_map}
}
\end{figure*}

In a three-planet system with two near-resonant pairs, neighboring resonances can both contribute significantly to the dynamics, giving rise to a chaotic region near the separatrix in the phase space. This has been demonstrated by a range of analytical and numerical studies \citep[e.g., ][]{Quillen2011, Rath2022_Criterion, Petit2020_path, Tamayo2021_criterion, Lammers2024_Instability}.

To investigate this phenomenon in AU Mic, V1298 Tau, and TOI-2076, we employed N-body simulations to map the chaotic regions. In this section, we demonstrate that configurations with circulating resonant angles exhibit significantly shorter Lyapunov timescales compared to their librating counterparts, confirming the more chaotic behavior in near-resonant systems.

\subsection{Simulation setup}
\label{subsec:sim-set}

We focus on three-planet systems, i.e. neighboring triplets, as they capture the essential dynamics of interacting resonant planets in a multi-planet system \citep{Rath2022_Criterion}. For systems with more than three planets, such as V1298~Tau, we analyze each adjacent triplet independently. The time step for every simulation is set to $P_1/20$, where $P_1$ is the initial orbital period of the innermost planet in a triplet. The integrator and the corresponding settings are the same as Section \ref{sec:stability}.

We construct a $150 \times 150$ grid in the period ratio space, i.e., $P_1/P_2$ vs. $P_3/P_2$, around the perfect period commensurability. At each grid point, we compute the Lyapunov timescale to assess how chaotic the orbit is. Planetary masses are fixed to the median of the TTV posterior distribution, while the raw eccentricities are initiated at $0.04$ to amplify the dynamical effects and clearly illustrate the onset of chaos. The initialization of the circulating and librating simulations is identical to Section \ref{sec:stability}. In the librating case, we only show only the grids where there is at least one pair in libration: defined by libration amplitudes smaller than 90$^{\circ}$ \citep{Millholland876}.

The Lyapunov timescale is estimated by fitting the slope of the Mean Exponential Growth factor of Nearby Orbits (MEGNO) indicator \citep{Cincotta2003_Phase}, as implemented by \texttt{REBOUND}. Following \citet{Rath2022_Criterion}, the simulation time is set to be 10 times the estimated Lyapunov timescale. If the linear fit to MEGNO yields a negative slope, which is often an artifact for long Lyapunov times, the simulation proceeds up to a maximum integration time, which is set to $10^6 P_2$. To improve the signal-to-noise ratio (SNR) of the resulting Lyapunov map, we perform 16 independent simulations per grid cell, each with randomized initial resonant angles, and adopt the median $T_{\rm Lyap}$ as the representative value for that grid cell.

\subsection{Map of the Lyapunov timescale}

Figure~\ref{fig:lyap_map} shows the Lyapunov timescale map for the AU Mic system as an illustrative example. The left and right panels show the circulating and librating configurations, respectively. The x-axis represents the period ratio of the inner pair, $P_1/P_2$, while y-axis represents the outer pair, $P_3/P_2$. The color scale represents the logarithmic Lyapunov timescale, with redder regions being chaotic and bluer regions being more regular or quasi-periodic. Black solid lines trace the nominal locations of zeroth-order 3BRs, with the shaded areas denoting their approximate analytical widths \citep[e.g., ][]{Petit2020_path}.

The detailed mechanisms that trigger chaos, as well as the locations of chaotic zones, have already been thoroughly investigated in previous studies \citep[e.g.,][]{Quillen2011, Petit2020_path, Tamayo2021_criterion, Rath2022_Criterion}. Very briefly, the situation is analogous to the classic double pendulum, chaos primarily develops when the middle planet of a triplet experiences resonant perturbations from both neighbors i.e. when $P_1/P_2$ and $P_3/P_2$ are both near resonant. In Figure~\ref{fig:lyap_map}, this corresponds to where $P_1/P_2=2/3$ (3:2 MMR) and $P_3/P_2=3/2$ (3:2 MMR) intersect. Most circulating orbits here are chaotic (Left panel), as both resonant terms exert comparable influence on the dynamics. However, we identify the following regions of the $P_1/P_2-P_3/P_2$ phase space to be quasi-periodic and long-term-stable particularly in librating configurations (right panel):

\begin{enumerate}
    \item {\bf Simultaneous 2BR and 3BR libration}: If both neighboring pairs are deep in librating 2BRs, the adjacent triplet often also librates in a 3BR \citep{Wang2024}. Dissipative processes such as disk migration can guide a system toward the stable equilibrium where both 2BRs and the 3BR librate simultaneously. Such a configuration is frequently produced in N-body disk migration simulations \citep[e.g.][]{Keller2025_Higher-Order} and are shown as yellow triangles in the right panel Figure~\ref{fig:lyap_map}.
    \item {\bf 3BR libration only}: Long-term stability can also be maintained by a librating 3BR even when neighboring pairs are not in 2BR. Here, libration of the 3BR angle prevents triple conjunctions and chaotic interactions. This configuration is consistent with many observed resonant chains (cyan points in Figure~\ref{fig:lyap_map}) that cluster along the nominal 3BR lines (black solid line). Systems in this category may arise from case (1) receding inner disk edge \citep[e.g. TRAPPIST-1,][]{Liu2017,Huang_Ormel,Hansen2024,Pichierri2024} or 2) tidal resonant repulsion \citep[e.g. Kepler-221,][]{Goldberg2021_Tidal}.
    \item {\bf Far from Resonances:} If planets are far from both 2BR and 3BR resonances (some observed examples are shown as lime squares in Figure~\ref{fig:lyap_map}), resonant interactions are weak, and the system typically remains dynamically stable and non-chaotic.
\end{enumerate}

Now, where does AU Mic reside on Figure~\ref{fig:lyap_map}? Since the TTV analysis indicates a circulating dynamical state, we focus on the left panel. The observed configuration of AU Mic resides in a mildly chaotic region, with a Lyapunov timescale on the order of $10^{3} - 10^{4}$ yr. V1298 Tau and TOI-2076 also occupy moderately chaotic zones. Because short Lyapunov timescales often correspond to limited stability lifetimes, we suspect that AU Mic, V1298 Tau, and TOI-2076 may be unstable. In Section \ref{sec:stability}, we directly test their long-term stability with N-body integrations.

\bibliography{sample631}{}
\bibliographystyle{aasjournal}

\end{document}